\documentstyle[12pt]{article}

\begin{document}
\mbox{ }
\rightline{UCT-TP-238/97}
\rightline{March 1997}
\vspace{3.0cm}
\begin{center}
\begin{Large}
{\bf  Vector Meson Dominance and $g_{\rho\pi\pi}$ at Finite Temperature
from QCD Sum Rules} 
\end{Large}

\vspace{1cm}

{\large {\bf C. A. Dominguez}$^{a}$},
{\large {\bf M. S. Fetea}$^{a}$}, and
{\large {\bf M. Loewe}$^{b}$}
\end{center}

\vspace{.5cm}

\begin{tabular}{lc}
$a$&Institute of Theoretical Physics and Astrophysics, University of Cape Town,\\
    &Rondebosch 7700, South Africa\\[.3cm]
$b$&Facultad de Fisica, Pontificia Universidad Catolica de Chile,\\
    &Casilla 306, Santiago 22, Chile
\end{tabular}
\vspace{1cm}

\begin{abstract}
A Finite Energy QCD sum rule at non-zero temperature is used to
determine the $q^{2}$- and the $T$ - dependence of the $\rho \pi \pi$
vertex function in the space-like region. 
A comparison with an independent QCD determination
of the  electromagnetic pion form factor $F_{\pi}$ at $T \neq 0$
indicates that Vector Meson Dominance holds to a very good approximation
at finite temperature. 
At the same time, analytical evidence for
deconfinement is obtained  from the result that  $g_{\rho \pi \pi}(q^{2},T)$
vanishes at the critical temperature $T_{c}$, independently of $q^{2}$.
Also, by extrapolating  the  $\rho \pi \pi$ form
factor to $q^{2} = 0$, it is found that the pion radius increases with
increasing $T$, and it diverges at $T=T_{c}$.
\end{abstract}
\noindent
\newpage
\setlength{\baselineskip}{1.5\baselineskip}
\noindent 
One of the popular reactions proposed for
probing the quark-gluon plasma is dilepton
production in  high energy heavy ion collisions \cite{QGP}. 
An important piece of information required to calculate the dilepton
production rates, in the hadronic phase, is the temperature variation
of the electromagnetic pion form factor, $F_{\pi}(q^{2})$. In 
these calculations it has been
usually assumed that Vector Meson Dominance (VMD) remains valid at
non-zero temperature and, with a few exceptions \cite{GRHOT}, that the
rho-meson mass and width are temperature independent. It has been
argued long ago, though, that all hadronic widths should increase with
increasing temperature, and presumably diverge at the critical temperature
for deconfinement \cite{GRHOT}-\cite{PIS}. This is expected to hold also 
for particles
which are (hadronically) stable at $T=0$, e.g. nucleons and pions. Actual
calculations in various frameworks do support such a scenario 
\cite{LS}-\cite{CAD2}.
In this sense, the imaginary part of a hadronic Green's function,
i.e. the width, may be viewed as a phenomenological signal for the occurrence
of deconfinement. In addition, a QCD sum rule determination of the pion
form factor at finite temperature clearly shows that it depends on $T$ in
such a way that it vanishes at the critical temperature, while the pion
radius diverges there \cite{FPIT}. This determination of $F_{\pi}(Q^{2},T)$
does not rely on any form of VMD, as it is based on the three-point function
associated to the electromagnetic current and two axial-vector currents,
thus projecting directly the electromagnetic pion form factor (in the
space-like region $q^{2} = - Q^{2} < 0$).

In this paper we study Finite Energy QCD sum rules (FESR) at $T\neq 0$
for the three-point function 
involving the rho-meson interpolating current plus two axial-vector
divergences. This allows us to determine the $Q^{2}$- and the $T$-
dependence of the $\rho \pi \pi$ coupling, as well as to
gauge the validity of VMD at finite temperature.  
We begin with the determination of $g_{\rho \pi \pi}(Q^{2})$
at zero temperature (for an earlier
analysis using Laplace sum rules see \cite{PN}), in order to establish
normalizations, as well as to check VMD here. This can be
accomplished by using $g_{\rho \pi \pi}(Q^{2})$  
determined from the sum rules, together with VMD, and comparing the
resulting pion form factor with the data. At finite temperature,
we can also compare it with the direct determination \cite{FPIT}, i.e.
with a theoretical result not relying on VMD. Since the latter does 
fit the data very well at $T=0$, 
we can adopt it as the benchmark $F_{\pi}(Q^{2},T)$ in the absence of
experimental data at $T \neq 0$. Agreement between the two
expressions could be taken as evidence in support of VMD at finite $T$.

We consider first the $T=0$ correlator
\begin{eqnarray}
\begin{array}{lcl}
\Pi_{\mu} (q) &=& i^{2} \int \int \; d^{4}x \; d^{4}y \;
e^{- i q \cdot y}  e^{ i p' \cdot x}< 0| \; 
T \; (j_{\pi}^{\dag} (x) \; J_{\mu}^{\rho} \; (y)  \; j_{\pi} (0) 
\;)|0>\\[.3cm]
   &=& \Pi_{1} (q^{2}) \; P_{\mu} + \Pi_{2} (q^{2}) \; q_{\mu} \; ,
\end{array}
\end{eqnarray}

where $J^{\rho}_{\mu}(y) = \frac{1}{2}
: [\bar{u}(y) \; \gamma_{\mu} \; u(y)- \;\bar{d}(y) \; \gamma_{\mu} d(y)]:$,
 $j_{\pi} (x) = (m_{u} \; + \; m_{d})\;
:\bar{d}(x) \; i  \gamma_{5} \; u(x):$, $q_{\mu} = (p' - p)_{\mu}$, 
and $P_{\mu} = (p' + p)_{\mu}$.
Calculating the imaginary part of the above three-point function in 
perturbative QCD to leading order in $\alpha_{s}$ and the quark masses 
gives the result
\begin{eqnarray}
\mbox{\mbox{Im}} \; \Pi_{\mu} |_{\mbox{QCD}} = \frac{3}{4}
 \frac{ \left( m_{u} + m_{d} \right)^{2}}
{ \left[ (s + s' + Q^{2})^{2} - 4ss' \right]^{3/2} }
\left[ - Q^{2} ss'  P_{\mu} + ss' (s - s') q_{\mu} \right] \; ,
\end{eqnarray}

where $s = p^{2}, s'= p'^{2}$, and  $Q^{2} = - q^{2} \ge 0$. The hadronic 
counterpart of this correlator may be obtained after saturation with the 
pion intermediate state, and using the current-field identity
\begin{eqnarray}
j_{\mu}^{a} = \frac{M_{\rho}^{2}}{f_{\rho}} \; \rho_{\mu}^{a}
\hspace{1.5cm} (a = 1,2,3),
\end{eqnarray}
where $j_{\mu}^{a}$ is the isospin current, $\rho_{\mu}^{a}$ is the
rho-meson field, and the experimental value of the coupling is
$f_{\rho} = 5.0 \pm 0.1$, as obtained from the decay rate of the rho-meson
into $e^{+} e^{-}$ \cite{PDG}.
The result for the hadronic spectral function, e.g. 
$\mbox{Im} \Pi_{1}$ is
\begin{eqnarray}
\mbox{Im} \; \Pi_{1} (s,s',Q^{2}) |_{\mbox{HAD}} &=& - 2 \; f_{\pi}^{2} 
\; \mu_{\pi}^{4}  \frac{M_{\rho}^{2}}{M_{\rho}^{2} + Q^{2}} 
\frac{g_{\rho \pi \pi} (Q^{2})}{f_{\rho}} 
\pi^{2} \delta(s - \mu_{\pi}^{2}) \; \delta(s' - \mu_{\pi}^{2}) \nonumber 
\\[.3cm]
&+& \theta(s - s_{0}) \;\theta(s' - s'_{0})\;
\mbox{\mbox{Im}} \; \Pi_{1}(s,s',Q^{2}) |_{\mbox{QCD}} \; ,
\end{eqnarray}

where $f_{\pi} = 93.2$ MeV, and $g_{\rho\pi\pi}(M^{2}_{\rho}) = 6.06 
\pm 0.03$ \cite{PDG}.
In principle, $g_{\rho\pi\pi}$ is a form factor, i.e. a function of $q^{2}$.
In the simpler version of VMD, i.e. single rho-meson dominance, this coupling
would be strictly constant. However, there are radial excitations of the
rho-meson (the $\rho$(1450), $\rho$(1700), etc.) which make a non-negligible 
contribution and turn the coupling into a form factor. For instance, naive 
VMD applied to the electromagnetic pion form factor predicts 
$g_{\rho \pi \pi}/f_{\rho} = 1$, while the experimental value is
20 \% higher. Hadronic models, such as e.g. the dual model \cite{CAD0},
account for
this difference by incorporating the rho-meson radial excitations; this gives
$g_{\rho \pi \pi}(0)/f_{\rho} = 1$, but
$g_{\rho \pi \pi}(M_{\rho}^{2})/f_{\rho} \simeq 1.2$, in agreement
with experiment. At the same time, naive VMD does not fit the data too well
in the space-like region \cite{CAD0}; much better fits are obtained by
allowing for a $q^{2}$-dependence of $g_{\rho\pi\pi}$.
In addition to the pion pole contribution in Eq.(4), there
are additional terms from the pionic radial excitations, the $a_{1}$
meson. etc.. However, we shall
include these in the hadronic continuum, which is supposed to be well 
approximated by the perturbative QCD expression Eq.(2), provided that the
thresholds $s_{0} \simeq s'_{0} > 1\; -\; 3 \; \mbox{GeV}^{2}$.

At this point one can invoke Cauchy's theorem which leads to Finite
Energy Sum Rules (FESR); the one of lowest dimension in the present case
is

\begin{equation}
 \int_{0}^{s_0} \; \int_{0}^{s'_0} \; \; \mbox{\mbox{Im}} \; 
\Pi_{1} (s,s')|_{\mbox{HAD}} \; ds \; ds' \;= \; 
\int_{0}^{s_0} \; \int_{0}^{s'_0} \; \; \mbox{\mbox{Im}} \; 
 \; \Pi_{1} (s,s')|_{\mbox{QCD}} \; ds \; ds' ,
\end{equation}

where $s_{0}$, $s'_{0}$, are the continuum thresholds, 
i.e. the onset of perturbative QCD.
After substitution of the QCD and hadronic
spectral functions  in the FESR one obtains

\begin{equation}
\frac{g_{\rho \pi \pi} (Q^{2})}{f_{\rho}} = \frac{3}{8 \pi^{2}} \;
\frac{f_{\pi}^{2}}{<\bar{q} q>^{2}} \; \frac{Q^{2}}{M_{\rho}^{2}} \;
(Q^{2} + M_{\rho}^{2}) \; I \; (Q^{2}) \; ,
\end{equation}

where
\begin{equation}
I(Q^{2}) = \frac{s_{0}}{16} \; \left( 3 + \frac{s_{0}}{Q^{2}} \right)
+ \frac{1}{8} \; \left( s_{0} + \frac{3}{4} Q^{2} \right) ln \; \left(
\frac{Q^{2}}{Q^{2} + 2 s_{0}} \right) \; ,
\end{equation}

and use was made of the Gell-Mann, Oakes and Renner (GMOR) relation 
\cite{GMOR}
\begin{equation}
f_{\pi}^{2} \; \mu_{\pi}^{2} = - (m_{u} + m_{d}) \; < \bar{q} q > \; .
\end{equation}
The result for $I(Q^{2})$ above was obtained after a double 
integration in the $s,s'$ plane. The region of integration is a triangle
in this plane, but use of other shapes doesn't introduce appreciable
differences in the numerical results for $g_{\rho \pi \pi}(Q^{2})$.
An advantage of using FESR as opposed to e.g. Laplace transform QCD
sum rules, is that the latter requires knowledge of the vacuum condensates
of all dimensions. In contrast, at most one condensate contributes
to a FESR of a given dimension. In  the present case, since the dimension 
of $\Pi_{1}$ is $d=2$, there are no condensates appearing in the lowest
dimensional FESR.
Invoking now Extended Vector Meson Dominance (EVMD), i.e. VMD but with 
allowance for a possible $Q^{2}$-dependence of $g_{\rho \pi \pi}$,
leads to a well known expression
for the electromagnetic pion form factor  

\begin{equation}
F_{\pi}(Q^{2})|_{\mbox{EVMD}} \; = \frac{M_{\rho}^{2}}{M_{\rho}^{2}+Q^{2}}
\frac{g_{\rho \pi \pi}(Q^{2})}{f_{\rho}} \;.
\end{equation}

Substituting the FESR result Eq.(6) into Eq.(9) gives

\begin{equation}
F_{\pi}(Q^{2})|_{\mbox{EVMD}}  \; = \frac{3}{8 \pi^{2}} \;
\frac{f_{\pi}^{2}}{<\bar{q} q>^{2}} \; Q^{2}
\left[ \frac{s_{0}}{16} \; \left( 3 + \frac{s_{0}}{Q^{2}} \right)
+ \frac{1}{8} \; \left( s_{0} + \frac{3}{4} Q^{2} \right) ln \; \left(
\frac{Q^{2}}{Q^{2} + 2 s_{0}} \right) \right]\;. 
\end{equation}

This result can be compared with that based on 
a three-point function involving the electromagnetic current and two
axial-vector currents \cite{PIONFF}, which projects the pion form factor
directly, and hence makes no use of VMD, i.e.

\begin{equation}
F_{\pi}(Q^{2}) \; = \frac{1}{16 \pi^{2}} \; \frac{1}{f_{\pi}^{2}}
\frac{s'_{0}}{(1 + Q^{2}/2 s'_{0})^{2}} \; .
\end{equation}

Although Eqs.(10) and (11) look structurally very different, they are
numerically very similar for $Q^{2} \ge 0.5 \;\mbox{GeV}^{2}$,
if $s_{0} \simeq \; 2.18 \; \mbox{GeV}^{2}$, and $s'_{0} \simeq 1 \;
\mbox{GeV}^{2}$. 
The latter value leads to
a very good fit of the data above $1\; \mbox{GeV}^{2}$ \cite{PIONFF}. 
On the other
hand, with $s_{0} \simeq 2.18\;\mbox{GeV}^{2}$, the 
extrapolation of Eq.(6) to
$Q^{2} = 0$ gives $g_{\rho \pi \pi}(0)/f_{\rho} \simeq 1$. It should be
stressed that the onset of the continuum need not be the same in the
two cases, as the correlators are different. A priori, all one knows is
that $s_{0}$ ($s'_{0}$) should roughly be in 
the region where the resonances loose
prominence, and the hadronic continuum takes over, i.e. somewhere in
the interval $1 - 3 \; \mbox{GeV}^{2}$. On the other hand, 
the extrapolation to
$Q^{2} = 0$ of any of the above results should not be taken too seriously,
as the Operator Product Expansion, the backbone of QCD sum rules, 
diverges at the origin.
The good numerical agreement between Eqs.(10) and (11), as shown in Fig.1,
may be seen as  a reflection of the validity of EVMD. In any case, our
main purpose here is not another fit to the data at $T=0$, but rather the
non-zero temperature behaviour of the $\rho\pi\pi$ form factor. Since it is
actually the ratio $g_{\rho\pi\pi}(Q^{2},T)/g_{\rho\pi\pi}(Q^{2},0)$ which
counts, modestly reasonable $T=0$ results are normally sufficient. 

Next, we reconsider the above FESR at finite temperature. Thermal 
corrections to $\Pi_{1} \; (q^{2})|_{\mbox{QCD}}$ can be calculated 
in the standard fashion \cite{BS}-\cite{CAD4}, and we 
find for the imaginary part

\begin{equation}
\mbox{Im} \; \Pi_{1} (s,s',Q^{2}, T) = \mbox{Im} \; \Pi_{1} (s,s',Q^{2},0)
\; F(s,s',Q^{2},T)\; ,
\end{equation}

where 
\begin{equation}
F(s,s',Q^{2},T)= 1 - n_{1} - n_{2} - n_{3} + n_{1} n_{2} + 
n_{1} n_{3} + n_{2} n_{3} \;, 
\end{equation}

\begin{equation}
n_{1} \equiv n_{2} \equiv n_{F} \; \left( \left| \frac{1}{2\; T}
\sqrt{\frac{x+y}{2}} \right| \right) \; ,
\end{equation}

\begin{equation}
n_{3} \equiv n_{F} \; \left( \left| \frac{Q^{2} + (x - y)/2}
{2\;T \; \sqrt{\frac{x+y}{2}}} \right| \right) \; ,
\end{equation}

$n_{F}(x) = (1 + e^{x})^{-1}$ is the Fermi thermal factor, and 
$x= s + s'$, $y= s - s'$.
On the hadronic side, both $f_{\pi}$ and $<\bar{q} q>$ will develop a
temperature dependence, and so will $s_{0}$. The latter follows from the
notion that as the resonance peaks in the spectral function become broader,
the onset of the continuum should shift towards threshold
\cite{CAD3}, \cite{Bar1}. The temperature behaviour of this asymptotic
freedom threshold has been obtained from the lowest dimension FESR 
associated to the two-point function involving the axial-vector currents 
\cite{CAD3}, \cite{Bar1}, with $f_{\pi} (T)$ as a known input (for a recent
updated discussion see \cite{MSF}). For temperatures not too close to
the critical temperature $T_{c}$, say $ T < 0.8\; T_{c}$, the following
scaling relation has been found to hold to a very good approximation
\cite{MSF}
\begin{equation}
\frac{f_{\pi}^{2}(T)}{f_{\pi}^{2}(0)} \simeq
\frac{<\bar{q} q>_{T}}{<\bar{q} q>} \simeq
\frac{s_{0}(T)}{s_{0}(0)}.
\end{equation}
We shall make use of this relation in the sequel, together with the
results of \cite{Bar2} for $f_{\pi}(T)$ and $<\bar{q} q>_{T}$, in the
chiral limit, as well as for $m_{q} \neq 0$. In addition, we shall
invoke the GMOR relation at finite $T$, which only gets modified at 
next to leading order in the quark masses \cite{MSF}.

The $T\neq$ 0 FESR now reads
\begin{equation}
\frac{g_{\rho \pi \pi} (Q^{2},T)}{f_{\rho}} = \frac{3}{8 \pi^{2}} \;
\frac{f_{\pi}^{2}(T)}{< \bar{q} q>^{2}_T} \; \frac{Q^{2}}{M_{\rho}^{2}} \;
(Q^{2} + M_{\rho}^{2}) \; I \; (Q^{2}, T) \; ,
\end{equation}
where
\begin{equation}
I(Q^{2}, T) = \frac{1}{8} 
\int_{0}^{s_{0}(T)} \; dx
\int_{- x}^{x} \; dy
\frac{(x^{2} - y^{2})}{(Q^{4} + 2 x Q^{2} + y^{2})^{3/2}}
F(x,y,Q^{2},T) \; ,
\end{equation}

and the integration in Eq.(18) must now be done numerically. The rho-meson 
mass was assumed to be temperature independent; the modest increase
of $M_{\rho}$ near $T_{c}$ as obtained from QCD sum rules \cite{REV},
and other methods \cite{CAD2}, does not change qualitatively the conclusions. 
The results for the ratio $g_{\rho\pi\pi}(Q^{2},T)/g_{\rho\pi\pi}(Q^{2},0)$
are shown in Fig.(2) for  $f_{\pi}(T)$ and $<\bar{q} q>_{T}$ in the
chiral limit (curve (a)), as well as for $m_{q} \neq 0$ (curve(b)).
Although $Q^{2} = 1 \mbox{GeV}^{2}$ was used in this figure, higher values of
$Q^{2}$ give similar results. Particularly, and importantly, the vanishing
of the ratio at or near the critical temperature is basically $Q^{2}$ -
independent. This provides analytical evidence for deconfinement. 
The good agreement between the pion form factor using
$g_{\rho\pi\pi}(Q^{2})$ plus VMD
and that obtained directly without invoking VMD,
persists at $T\neq0$.
An extrapolation of these results to $Q^{2} = 0$ allows for a
determination of the $\rho\pi\pi$ root mean squared radius. Although this
is divergent at any temperature because of mass singularities, the ratio
$<r^{2}>(T)/<r^{2}>(0)$ is well defined; it increases with increasing
$T$ until the critical temperature where it becomes infinte, thus signalling
deconfinement. The temperature behaviour of this ratio is shown in Fig. 3.\\

To conclude, we obtained direct evidence supporting Extended VMD at $T=0$
from a comparison of the $\rho\pi\pi$ vertex function determined from
QCD sum rules and the experimental data on the electromagnetic pion form
factor. At finite temperature, and in the absence of data, we compared
our determination with a benchmark pion form factor obtained from 
independent QCD sum rules and not
using VMD. The close agreement found between the two expressions may be
interpreted as supportive of (Extended) VMD at $T\neq 0$. At the same
time, our determination provides additional analytical evidence for
deconfinement, as   $g_{\rho\pi\pi}(Q^{2},T)$ decreases with $T$, 
vanishing at the critical temperature, while the ratio of the 
$\rho\pi\pi$ root mean square radii $<r^{2}>(T)/<r^{2}>(0)$ increases
with temperature and becomes infinite at $T=T_{c}$.\\

{\bf Acknowledgements} The work of (CAD) and (MSF) has been supported in
part by the FRD (South Africa), and that of (ML) by Fondecyt (Chile)
under grant No.1950797, and Fundacion Andes under grant No.C-12999/2.\\

\section*{Figure Captions}
\begin{description}
\item[Figure 1:] The electromagnetic pion form factor at $T=0$, determined
from the QCD-FESR, Eq.(11), i.e. without invoking VMD
(dashed curve (a)), compared with the result of the independent QCD-FESR
for $g_{\rho \pi \pi}$ plus VMD, Eq.(10) (solid curve (b)). Experimental
data is from \cite{FPIEXP}.
\item[Figure 2:] The ratio of the $\rho \pi \pi$ form factors at finite
and zero temperatures as a function of $T/T_{c}$. Curve (a) uses $f_{\pi}
(T)$, $<\bar{q} q>(T)$, and $s_{0}(T)$ in the chiral limit, and curve
(b) away from this limit ($m_{q} \neq 0$). Both curves are for $Q^{2}
= 1 \mbox{GeV}^{2}$, while other values of $Q^{2}$ give essentially the same
ratios.
\item[Figure 3:] The ratio of the (strong) pion radius at finite and
at zero temperatures as a function of $T/T_{c}$.
\end{description}
\end{document}